\documentclass{article}

\usepackage[numbers]{natbib}

\usepackage[final]{neurips_2020_ml4ps}

\usepackage[utf8]{inputenc} 
\usepackage[T1]{fontenc}    
\usepackage{hyperref}       
\usepackage{url}            
\usepackage{booktabs}       
\usepackage{amsfonts}       
\usepackage{nicefrac}       
\usepackage{microtype}      
\usepackage{graphicx}
\usepackage{bm}
\usepackage{amsmath,amsfonts,amssymb}

\title{Probabilistic neural network-based reduced-order surrogate for fluid flows}

\author{
  Kai Fukami$^{1,2}$, Romit Maulik$^{3}$, Nesar Ramachandra$^{4}$, Koji Fukagata$^{2}$, Kunihiko Taira$^{1}$\\
  1. Mechanical and Aerospace Engineering, University of California, Los Angeles \\
  2. Mechanical Engineering, Keio University\\
  3. Argonne Leadership Computing Facility, Argonne National Laboratory\\
  4. High Energy Physics Division, Argonne National Laboratory\\
  Corresponding author: kfukami1@g.ucla.edu
  }

\newcommand\redsout{\bgroup\markoverwith{\textcolor{red}{\rule[0.5ex]{2pt}{0.4pt}}}\ULon}

\begin{document}

\maketitle
\vspace{-10mm}

\begin{figure}[h]
    \centering
    \includegraphics[width=0.8\textwidth]{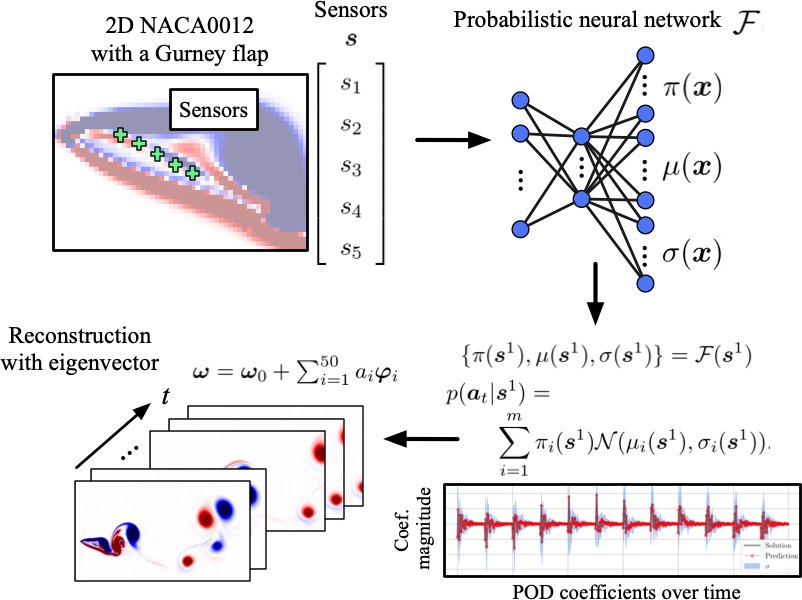}
    \caption{{Overview of probabilistic neural network-based reduced-order modeling for fluid flows.}}
    \label{fig1}
\end{figure}

\vspace{-3mm}
\begin{abstract}
\vspace{-3mm}
In recent years, there have been a surge in applications of neural networks (NNs) in physical sciences. 
Although various algorithmic advances have been proposed, there are, thus far, limited number of studies that assess the interpretability of neural networks. 
This has contributed to the hasty characterization of most NN methods as ``black boxes'' and hindering wider acceptance of more powerful machine learning algorithms for physics. 
In an effort to address such issues in fluid flow modeling, we use a probabilistic neural network (PNN) that provide confidence intervals for its predictions in a computationally effective manner.
The model is first assessed considering the estimation of proper orthogonal decomposition (POD) coefficients from local sensor measurements of solution of the shallow water equation.
We find that the present model outperforms a well-known linear method with regard to estimation.
This model is then applied to the estimation of the temporal evolution of POD coefficients with considering the wake of a NACA0012 airfoil with a Gurney flap and the NOAA sea surface temperature.
The present model can accurately estimate the POD coefficients over time in addition to providing confidence intervals thereby quantifying the uncertainty in the output given a particular training data set. 

\end{abstract}

\section{Introduction}

The use of neural networks (NNs) as universal approximators has recently attracted great attention in various problems in physical sciences.
Despite the recent enthusiasm, there are limited studies of interpretable NN methods, which have led to the characterization of neural networks as ``black-boxes" which do not provide feedback from a modeling strategy. 
A decisive reason for this is the specification of modeling tasks from the deterministic regression point-of-view: $L_2$ norms are well-understood and widely utilized in linear least-squares applications. However, this limits the amount of information one can extract from a modeling effort. In contrast, Bayesian models, for e.g., variational inference \cite{blundell2015} and Gaussian process approximations \cite{damianou2019gp} offer significant flexibility, and may be utilized to assess model and data quality as well as quantify uncertainty under one umbrella.

In the present paper, we introduce the use of probabilistic neural network (PNN) that assumes a generative distribution for the target data \cite{bishop1994mixture}.  The PNN is able to estimate the expected value of the targets while also predicting a confidence interval for each estimation. For our presentation, we consider the construction of reduced-order models for fluid dynamics.

\section{Methods}
\subsection{Probabilistic neural network (PNN)}

Neural networks $\cal F$, which output ${\bm y}$ from inputs $\bm x$, are generally trained to obtain optimized weights ${\bm w}$ by minimizing a cost function $E$ such that ${\bm w}={\rm argmin}_{\bm w}[E({\bm y},{\cal F}({\bm x};{\bm w}))]$.
The output obtained through this process has no information about the estimation uncertainty. To obtain the probability distribution of estimations $p({\bm y}_p|{\bm x})$, we consider the use of a PNN.

Instead of the general map for an NN, $\mathcal{F}: {\bm x} \rightarrow {\bm y}({\bm x})$, the mapping of a PNN can be written as $\mathcal{F}: {\bm x} \rightarrow (\pi_1, \mu_1, \sigma_1, \pi_2, \mu_2, \sigma_2, ...,  \pi_N, \mu_N, \sigma_N)$, where $\pi_i$ is the mixing probability for each Gaussian component satisfying the condition $\sum_{i=1}^{m}\pi_i = 1$, the mean $\mu$ and standard deviation $\sigma$ parametrize a Gaussian probability distribution function $\mathcal{N}(\mu,\sigma)$. 
Therefore, the distribution function in this model is a linear combination of several Gaussian components,
\begin{equation}\label{eq:GMM}
    p({\bm y} | {\bm x}) = \sum_{i=1}^{m}\pi_i ({\bm x}) \mathcal{N}(\mu_i ({\bm x}),\sigma_i ({\bm x})).
\end{equation}
The value of $m$ is generally pre-specified based on the expectation of posterior distribution.
Utilizing the mixture of Gaussians in the manner above, we can expect to be able to handle a complex probability distribution for the output of PNN $p({\bm y}_p|{\bm x})$. Because our PNN attempts to output a distribution of the estimation $p({\bm y}_p|{\bm x})$ instead of a target variable of ${\bm y}$ directly, care should be taken to choose the loss function so that the full distribution of the estimation can be utilized.  In the present study, the cost function $E$ for the PNN ${\mathcal F}({\bm x};{\bm w})$ is given in terms of average log-likelihood $\mathcal{L}$ such that
\begin{align}\label{eq:negloglike}
\begin{gathered}
{\bm w}={\rm argmin}_{\bm w}[\mathcal{E}],\quad \text{where} \quad \mathcal{E} \equiv -\log \mathcal{L} = -\sum_{k=1}^{K} p({y}_{k,p}|{x_k}) \log p({y}_{k,t})
\end{gathered}
\end{align}
with $k$ indicates each data point in the training data and $K$ denotes a number of training samples. 
The term $p({y}_{p,k}|{x_k})$ in the error function is evaluated for each data point using the output of the network given in equation \ref{eq:GMM}.  Note that the likelihood maximizing model such as the present model is equivalent to minimizing the cross entropy $H(p({\bm y}_p|{\bm x}), p({\bm y}_t) )$, which encourages us to apply the model to a wide variety of problems.

\subsection{PNN-based reduced-order model with proper orthogonal decomposition}

In the present study, we demonstrate the capability of the PNN for predicting the temporal evolution of coefficients for proper orthogonal decomposition (POD) from the state at the first instantaneous field, as shown in figure \ref{fig1}.
Using the POD basis $\bm \varphi$ and the POD coefficients, the state $\bm q$ can be decomposed as ${\bm q}={\bm q}_0+\sum_{i=1}^{M}{a_i}{\bm \varphi}_i$, where ${\bm q}_0$ is the temporal average of the state and $M$ denotes the number of POD modes.
The present PNN attempts to predict the POD coefficients over $n$ time steps ${\bm a}_t=[{\bm a}^1,{\bm a}^2,...,{\bm a}^{n}]$, where ${\bm a}^{\iota}=[a^{\iota}_1,a^{\iota}_2,...,a^{\iota}_{M}]$, from the state (e.g., sensor information) at the first instantaneous field ${\bm s}^1$,
\begin{align}
\begin{gathered}
\{\pi({\bm s}^1),\mu({\bm s}^1),\sigma({\bm s}^1)\}={\cal F}({\bm s}^1),~~~p({\bm a}_t | {\bm s}^1) = \sum_{i=1}^{m}\pi_i ({\bm s}^1) \mathcal{N}(\mu_i ({\bm s}^1),\sigma_i ({\bm s}^1)).
\end{gathered}
\end{align}
The present problem setting can be regarded as a reduced-order surrogate for high-dimensional systems since the original dynamics can be approximated in both space and time by combining the estimated POD coefficients and the POD basis.
In this study, we will assess the proposed modeling framework for the wake estimation of a NACA0012 airfoil with a Gurney flap and the flow reconstruction of the sea surface temperature (from the NOAA Optimum Interpolation dataset).

\section{Results}
\subsection{Comparison with benchmark linear method}

Let us first consider the estimation of POD coefficients ${\bm a}$ from local sensor measurements ${\bm s}$ at a given instant in time, by using the two-dimensional inviscid shallow water equations which are a prototypical system for geophysical flows. 
The problem setting can be mathematically expressed as
\begin{align}
\begin{gathered}
\{\pi({[{\bm s},{q_t}]}),\mu([{\bm s},{q_t}]),\sigma([{\bm s},{q_t}])\}={\cal F}([{\bm s},{q_t}]),~~~p({\bm a} | [{\bm s},{q_t}]) = \sum_{i=1}^{m}\pi_i ([{\bm s},{q_t}]) \mathcal{N}(\mu_i ([{\bm s},{q_t}]),\sigma_i ([{\bm s},{q_t}])).
\end{gathered}
\end{align}
As shown, the time stamp, $q_t$, which indicates the progress to the final time of the evolution is also utilized as an input attribute in addition to the sensor measurements.
\begin{figure}
    \centering
    \includegraphics[width=0.9\textwidth]{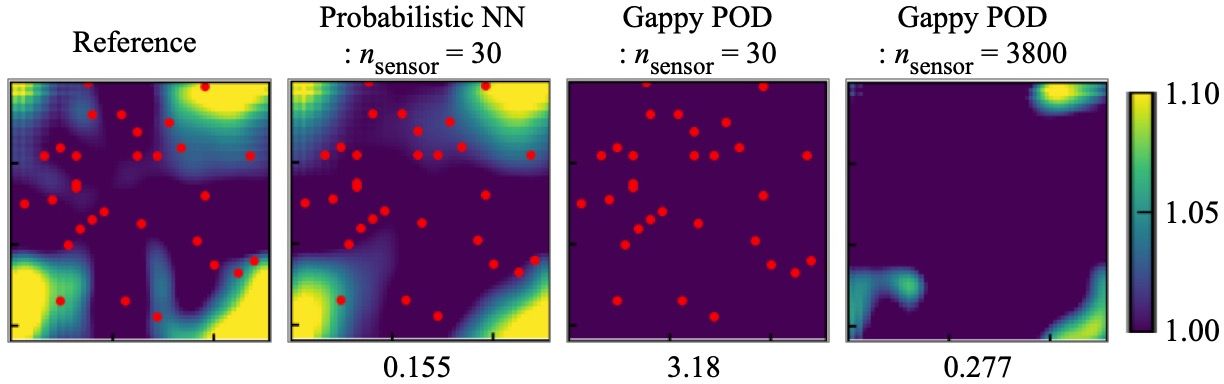}
    \caption{A comparison between PNN and Gappy POD for POD coefficient reconstructions. The values underneath the contours indicate the normalized $L_1$ error norms. Sensor locations (shown in red) are not provided for the validation with 3800 sensors for visual clarity.}
    \label{fig0}
\end{figure}
We here compare the PNN and a well-known linear reconstruction method --- Gappy POD \cite{everson1995karhunen} for flow reconstruction, as presented in figure \ref{fig0}. 
The present model outperforms Gappy POD at the same number of sensors $n_{\rm sensor}=30$. Even at $n_{\rm sensor}=3800$, the PNN exhibits a significant advantage, as observed through an $L_1$ norm comparison. 
Note that the present model can provide confidence intervals for coefficient estimation as well (demonstrated in the following sections), in comparison to Gappy POD which is determinstic.

\subsection{NACA0012 airfoil with a Gurney flap}

\begin{figure}
    \centering
    \includegraphics[width=0.7\textwidth]{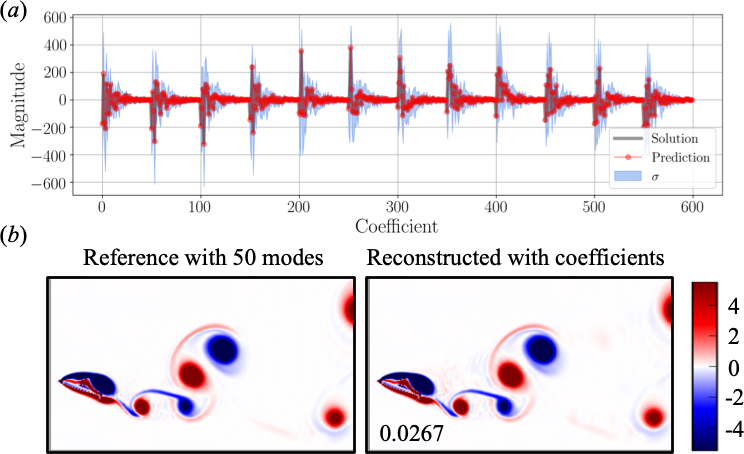}
    \caption{Application of probabilistic neural network to the wake of NACA0012 airfoil with a Gurney flap. $(a)$ Estimation of POD coefficients over a half of vortex shedding period. $(b)$ Final instantaneous vorticity snapshots reconstructed by the estimated coefficients. The value inside the contour indicates the $L_2$ error norm of reconstructed field with the estimated coefficients.}
    \label{fig2}
\end{figure}

Next, let us consider the wake behind a NACA0012 airfoil with a Gurney flap. 
The training data set is prepared using two-dimensional direct numerical simulation at a chord based Reynolds number of ${\rm Re}_c =1000$ \cite{GMTA2018}. 
We use 100 snapshots collected over approximately 4 periods for the training of PNN. 
For the example of NACA0012 wake, the PNN tries to predict the temporal evolution of fifty POD coefficients for the vorticity field ${\bm a}^{\iota}=[a^{\iota}_1,a^{\iota}_2, a^{\iota}_{3},...,a^{\iota}_{50}]$ over 12 time steps ${\bm a}_t=[{\bm a}^1,{\bm a}^2,...,{\bm a}^{100}]$ corresponding to approximately a half of vortex shedding period, from the five vorticity sensor measurements on surface of an airfoil at the first instantaneous field ${\bm s}^1$. 
Hence, the dimensions of input and output are 5 and 600, respectively.

Our investigation for the NACA0012 with a Gurney flap is summarized in Figure \ref{fig2}.
The PNN can accurately estimate the temporal evolution for the POD coefficients while showing its confidence interval as shown in figure \ref{fig2}$(a)$, which corresponds to the normalized $L_2$ error norm for the POD coefficients $\epsilon_c=||{\bm a}_{\rm Ref}-{\bm a}_{\rm PNN}||_2/||{\bm a}_{\rm Ref}||_2$ of 0.145.
By combining to the POD basis, the high-dimensional flow can also represented well with the normalized $L_2$ error norm for the reconstructed fields $\epsilon_f=||{\omega}_{\rm Ref}-{\omega}_{\rm PNN}||_2/||{\omega}_{\rm Ref}||_2$ of 0.0267, as presented in figure \ref{fig2}$(b)$.

\subsection{NOAA sea surface temperature}

We also consider the NOAA sea surface temperature data set for field reconstruction from limited sensor measurements.
The spatial resolution of data set is $360\times180$ based on a one degree grid and is obtained from satellite and ship observations.
Training data set is prepared from 20 years of data (1040 snapshots spanning 1981 to 2001), while the test data set is obtained from 874 snapshots from years 2001 to 2018.
The problem setting here is inspired by Callaham et al. \cite{CMB2019} who capitalized on sparse representations to reconstruct fluid flow fields from local sensors. 
For the task with the sea surface temperature, the PNN predicts the temporal evolution of 4 POD coefficients ${\bm a}^{\iota}=[a^{\iota}_1,a^{\iota}_2, a^{\iota}_{3},a^{\iota}_{4}]$ over 100 weeks ${\bm a}_t=[{\bm a}^1,{\bm a}^2,...,{\bm a}^{100}]$ from the local 30 sensor measurements on the first week snapshot ${\bm s}^1$.
Following Callaham et al. \cite{CMB2019}, the input sensors are chosen randomly from the region between $50{^\circ}$ S and $50{^\circ}$ N. 

\begin{figure}
    \centering
    \includegraphics[width=0.70\textwidth]{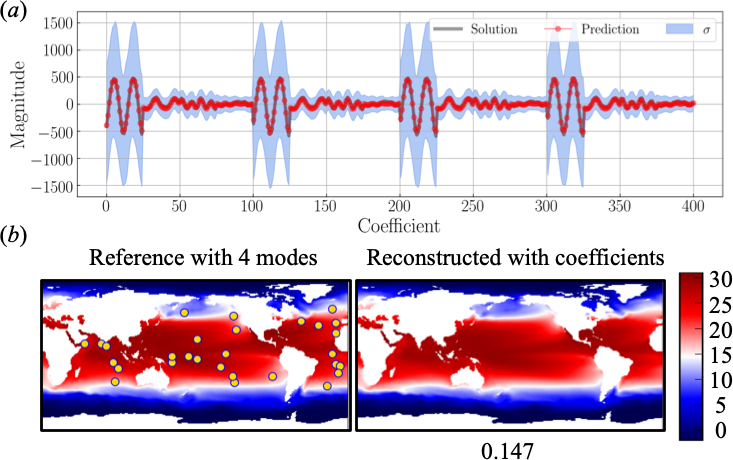}
    \caption{Application of probabilistic neural network to the NOAA sea surface temperature. $(a)$ Estimation of POD coefficients over 100 weeks. $(b)$ Final instantaneous snapshots reconstructed by the estimated coefficients. The reference contour includes input sensor locations. The value underneath the contour indicates the $L_2$ error norm of reconstructed field with the estimated coefficients.}
    \label{fig3}
\end{figure}

The results for NOAA sea surface temperature are presented in figure \ref{fig3}. 
As shown, the temporal trend of POD coefficients can be captured well using PNN.
What is striking here is that the estimated uncertainty is larger than that of NACA0012 example, which is likely caused by the noisy nature of real-world data. This drives home an important point about the quality of training data for neural network optimization --- probabilistic methods offer insight into the nature of an experimentally generated data set as well.
Since estimation of POD coefficients works well, the reconstruction for the temperature field is also in good agreement with the reference data.

\section{Concluding remarks}

This article introduced the use of probabilistic neural networks (PNN) to quantify uncertainties within a surrogate model for fluid flow reconstruction and forecasting applications.
The advantage against the conventional linear method was first demonstrated with a shallow water equation.
We then assessed the framework on the wake estimation of a NACA0012 airfoil with a Gurney flap and the NOAA sea surface temperature data set. 
The present PNN showed promising potential as a reduced-order surrogate with uncertainty estimates of the targets conditioned on the inputs and the training data.
Although not shown here, the confidence intervals obtained by the present modeling strategy may also be utilized additional sensor placements in fluid flow data recovery tasks \cite{MFRFT2020}.
These observations enable us to expect various extensions of PNN to a wide spread of engineering applications.



\small


\newpage
\bibliographystyle{unsrt}  
\bibliography{main}

\section*{Broader impact}
The considerable positive impact of this work is that the devised reduced-order surrogate can reduce the computational cost for high-fidelity simulations over the course of a computational campaign significantly. The quantified uncertainty of the surrogate allows for greater insight into training data fidelity and may thus be utilized for experimental cost reduction via coupling of optimal data generation strategies with the machine learning. This would also engender considerable cost reduction in the experimental phase of several engineering and geophysical applications. There is no negative impact associated with this work. 

\section*{Acknowledgements}
This material is based upon work supported by the U.S. Department of Energy (DOE), Office of Science, Office of Advanced Scientific Computing Research, under Contract~DE-AC02-06CH11357. This research was funded in part and used resources of the Argonne Leadership Computing Facility, which is a DOE Office of Science User Facility supported under Contract DE-AC02-06CH11357. R.M. acknowledges support from the ALCF Margaret Butler Fellowship. K. Fukami and K. Fukagata thank the support from Japan Society for the Promotion of Science (KAKENHI grant number: 18H03758). K.T. acknowledges the generous support from the US Army Research Office (grant number: W911NF-17-1-0118) and US Air Force Office of Scientific Research (grant number: FA9550-16-1-0650). The authors acknowledge Dr. M. Gopalakrishnan Meena (Oak Ridge National Laboratory) for sharing his DNS data.  

\end{document}